\documentclass[aps,
		prd,
		reprint,
		onecolumn,
		superscriptaddress,
		shortbibliography,
		nofootinbib,
		floatfix,
		notitlepage
		]{revtex4-1}
\pdfoutput=1
\usepackage{amsmath,amssymb,amsfonts}

\usepackage{mathletters}

\usepackage[dvipsnames]{xcolor}

\usepackage{bm}
\usepackage{amsmath}
\usepackage{amssymb}
\usepackage{graphicx}
\usepackage{hyperref}
\usepackage{cleveref}
\usepackage{slashed}
\usepackage{amsfonts}
\usepackage{amsbsy}
\usepackage{color}
\usepackage{array}
\usepackage{dcolumn}
\usepackage{tensor}
\usepackage{braket}
\usepackage{mathrsfs}
\usepackage{tikz}
\usepackage{verbatim}

\newcommand{\be}{\begin{equation}}
\newcommand{\ee}{\end{equation}}
\newcommand{\bi}{\begin{itemize}}
\newcommand{\ei}{\end{itemize}}
\newcommand{\LCm}{{\scriptscriptstyle -}} 
\newcommand{\LCp}{{\scriptscriptstyle +}}
\newcommand{\LCpm}{{\scriptscriptstyle \pm}}
\newcommand{\LCmp}{{\scriptscriptstyle \mp}}

\newcommand{\LCperp}{{\scriptscriptstyle \perp}}

\begin{document}

\title{Particle-beam scattering from strong-field QED}

\author{Tim Adamo}
\email{t.adamo@ed.ac.uk}
\affiliation{School of Mathematics and Maxwell Institute for Mathematical Sciences, University of Edinburgh, EH9 3FD, UK}

\author{Anton Ilderton}
\email{anton.ilderton@ed.ac.uk}

\affiliation{Higgs Centre, School of Physics and Astronomy, University of Edinburgh, EH9 3FD, UK}

\author{Alexander J.~MacLeod}
\email{alexander.macleod@plymouth.ac.uk}
\affiliation{Centre for Mathematical Sciences, University of Plymouth, PL4 8AA, UK}

\begin{abstract}
We consider the scattering of probe particles on an ultra-boosted beam of charge, in the case that the fields of the beam are strong and must be treated non-perturbatively. 
We show that the fields of the ultra-boosted beam act as stochastic plane waves -- scattering amplitudes (of elastic scattering, nonlinear Compton and nonlinear Breit-Wheeler) are obtained without approximation by averaging plane wave scattering amplitudes over all possible plane wave parameters. 
The relevant plane waves are ultra-short and, as such, scattering on ultra-boosted beams does not exhibit the conjectured strong-field behaviour of QED based on the locally constant field approximation.
\end{abstract}

\maketitle

\section{Introduction \label{sec:Intro}}

A strong background field is characterised by a coupling to charged particles which is larger than unity. 
Interactions with such a background thus require non-perturbative methods. 
One terrestrial source of strong \textit{electromagnetic} fields is modern high-power lasers~\cite{ELI,CORELS}, which offer prospects for both theoretical and experimental investigations of quantum field theory at strong coupling. One of the goals of current and upcoming laser facilities is indeed to probe `strong-field QED' {effects~\cite{Abramowicz:2021zja,E320}; that these can be measured} in laser-particle collisions has recently been demonstrated in experiments on radiation reaction~\cite{Cole:2017zca,Poder:2017dpw}. 

Beam-beam interactions have also been suggested as a method to probe the strong-field regime of QED~\cite{DelGaudio:2018lfm,Yakimenko:2018kih}, and ultraperipheral heavy ions collisions have seen success in the measurement of light-by-light scattering~\cite{ATLAS:2017fur,ATLAS:2019azn,CMS:2018erd} and pair production~\cite{STAR}, highlighting the feasibility of exploring field-induced phenomena using beam-beam collisions. For high total charge, or small impact parameter, the electromagnetic fields of one beam (of electrons, positrons, or ions) as seen by the other will be strong, and require a non-perturbative treatment. 

The theoretical framework for strong-field QED calculations is well-established. Amplitudes for processes are calculated in the Furry expansion~\cite{Furry:1951zz}, that is, background field perturbation theory. 
The strong field in question is treated without approximation as a fixed background, and scattering processes on the background are calculated in perturbation theory. {While studies of laser-particle collisions have traditionally followed this approach (usually modelling the laser fields as plane waves) strong field QED in beam-beam interactions has been studied numerically using particle-in-cell simulations and models based on the (locally) constant field approximation~\cite{DelGaudio:2018lfm,Yakimenko:2018kih}. This assumes that the strong field varies slowly over scales relevant for processes to occur~\cite{Ritus:1985}. }
However, if the physical situation of interest is the collision of heavily boosted beams, particles or ions, then the fields of one bunch as seen by the other will be very strongly Lorentz contracted and hence switch \textit{rapidly} on and off.
It is thus natural (not least given the well-studied shortcomings of the locally constant field approximation~\cite{Harvey:2014qla,DiPiazza:2017raw,Ilderton:2018nws,DiPiazza:2018bfu}) to investigate other methods for studying strong-field effects in beam-beam collisions. 
Our aim here is to take a first step in this direction,  considering the interaction between single electrons or photons with ultrarelativistic beams of electrons or positrons with high total charge, implying strong fields.

The standard starting point for calculations in strong-field QED is the construction of \textit{exact} solutions to the Dirac equation in the chosen background~\cite{volkov35,Bagrov:1990xp,Heinzl:2017zsr}. 
These provide the asymptotic fermion wavefunctions and propagators for scattering amplitudes calculated in the Furry picture. 
The construction and properties of the solutions describing an electron in the (strong) field of a heavily boosted bunch of charge  will be the focus of the first part of this paper. 
Here we capitalise on known models and methods in high-energy physics and gravity, where exact solutions of the Maxwell and Einstein equations corresponding to beams of massless particles were found over 50 years ago by Bonnor~\cite{Bonnor:1969rb,Bonnor:1969gr}. 
When the beam radius shrinks to zero, these solutions become \emph{shockwaves}: exact solutions whose source is localised on the lightfront as well as in the transverse plane. 
Alternatively, such shockwave solutions can be obtained, in electromagnetism, by ultraboosting the Coulomb field of a point source~\cite{Jackiw:1991ck}. 
In general relativity, one boosts instead the Schwarzschild solution~\cite{Aichelburg:1970dh}.

Shockwaves have been used extensively in the literature as a tool to describe $2\to2$ scattering in the eikonal regime, where the momentum transfer is small relative to the energy of the scattering process. 
In this eikonal limit, dominant contributions to the scattering amplitude are given by ladder diagrams with exchanges of the highest relevant spin~\cite{tHooft:1987vrq} (e.g., photon exchanges in QED or graviton exchanges in any gravitationally-coupled theory) which, under favourable circumstances (e.g., \cite{Tiktopoulos:1971hi,Eichten:1971kd}), can be resummed so that the amplitude exponentiates in impact parameter space and is entirely determined by an eikonal phase factor, c.f.~\cite{Levy:1969cr}. Neglecting masses (consistent with the high-energy regime), the eikonal amplitude is reproduced by semi-classical $1\to1$ scattering of a massless probe in a shockwave background~\cite{Dray:1984ha,tHooft:1987vrq,Jackiw:1991ck,Kabat:1992tb}. More generally, shockwaves play an important role in describing high-energy hadron collisions through the Balitsky-JIMWLK and colour-glass condensate formalisms~\cite{Balitsky:2001gj,Gelis:2010nm,Kovchegov:2012mbw,Caron-Huot:2013fea}, and as probes of causality, universality and `transplanckian scattering' -- where the centre of mass enery exceeds the Planck scale -- in quantum gravity~\cite{Amati:1987wq,Amati:1988tn,Horowitz:1989bv,Verlinde:1991iu,Giddings:2004xy,Giddings:2010pp,Ciafaloni:2014esa,Camanho:2014apa,Kologlu:2019bco}.

\medskip

In any case, beam and shockwave solutions provide classical backgrounds which, in the context of strong field QED, can be seen as approximations to the strong electromagnetic fields of an accelerated bunch of charge. To study scattering of particles on the bunch one has to formulate, as indicated above, the Furry expansion in the beam/shockwave background.
Quantum field theory in a shockwave background has been studied in the literature, including by Balitsky in the context of rapidity evolution of colour dipoles in high-energy QCD~\cite{Balitsky:1995ub,Balitsky:2004rr,Balitsky:2005we}, and by Lodone and Rychkov in the context of gravitational transplanckian scattering~\cite{Lodone:2009qe}. 
Remarkably, strong-field QED in beam or shockwave backgrounds has not been systematically studied\footnote{An exception is for $1\to1$ scattering in an electromagnetic shockwave, which has been shown to reproduce the standard eikonal amplitude of QED~\cite{Jackiw:1991ck}.}: we begin this programme here.

This paper is organised as follows. 
In \cref{sec:Classical} we give the fields of heavily boosted particles and beams, and solve the classical equations of particle motion in those fields. 
In \cref{sec:Quantum} we turn to the quantum theory, solving the Dirac equation in our chosen backgrounds. 
We use the solutions to calculate the simplest, but non-trivial ($1 \to 1$) tree level scattering amplitude and compare with the literature. 
In \cref{sec:3Point} we make a connection with laser-particle scattering; we show that, despite the very different physical situations, the exact solutions of the Dirac equation can be written without approximation in terms of the Volkov solutions of the Dirac equation in a background plane wave~\cite{volkov35}.
We then turn to the calculation of amplitudes, probabilities and cross-sections for the three-point processes most relevant for upcoming experiments, namely nonlinear Compton scattering and nonlinear Breit-Wheeler pair production. {We examine the cross-sections, exploiting connections with the Volkov wavefunctions, and paying particular attention to field-strength dependence and effects which cannot be captured by the locally constant field approximation.  We conclude in \cref{sec:Summary}.}

\section{Classical ultraboosted beams and particle dynamics~\label{sec:Classical}}

An ultraboosted beam is a highly boosted bunch of charge such that the self-interactions in the beam can be neglected~\cite{Chen:1987zg}. 
The electromagnetic field of the bunch, as seen by probe particles, is simply the superimposed boosted Coulomb fields of the charges. In this section, we review the electromagnetic fields of such ultraboosted beams (including the shockwave limit, where the beam's transverse radius shrinks to zero), as well as the classical dynamics of charged particles moving in the fields of the beam.

\subsection{The fields of ultraboosted beams~\label{sec:ShockPotential}}

Consider a highly boosted beam of charge travelling in the $z$-direction; it is convenient in the ultrarelativistic limit to use the usual lightfront coordinates~\cite{Bakker:2013cea} $x^\LCm = (t - z)/\sqrt{2}$ (lightfront time), $x^\LCp = (t + z)/\sqrt{2}$ (longitudinal to the beam) and $x^\LCperp=(x,y)$ (transverse to the beam). 
For momenta we have $p_\LCpm =(p_0 \pm p_3)/\sqrt{2} = p^\LCmp$, $p_\LCperp=(p_1,p_2)$. 
We introduce the vector $n_\mu$ defined by $n\cdot x = x^\LCm$, and define $r:=|x^\LCperp|$. 
In the ultrarelativistic limit, the potential of a bunch of total charge $Q$ and transverse radius $r_0$ may be chosen as~\cite{Bonnor:1969rb}
\begin{align}\label{eqn:BrinkmannGauge}
    A_{\mu}
    =
    -
    n_{\mu}
    \delta(n \cdot x)
    \Phi(x^{\lcperp})
    \;,
    \qquad
    \Phi(x^{\lcperp})
    =
    \frac{Q}{4\pi}\left\{
        \begin{array}{l  r}
            1 + \log(r^{2}/r_{0}^{2}) & r \ge r_{0}
            \\
            r^{2}/r_{0}^{2} & r \le r_{0}
        \end{array}
    \right.
    \,.
\end{align}
The delta-function of lightfront time arises directly from the classical current of an ultraboosted particle, and the electromagnetic fields of the beam, 
\begin{align}\label{eqn:FieldStrength}
    F_{\mu\nu} 
    = 
    \delta(n\cdot x)
    \big(
        n_\mu \partial_\nu 
        - 
        n_\nu \partial_\mu 
    \big)  
    \Phi(x^{\lcperp})
    \;,
\end{align}
are pancaked into a disc, moving at the speed of light in the $z$-direction. 
The fields are thus localised in time but not in transverse space, where they fall off as $1/r$.

If we take the beam radius to be negligible, or if the source is a single ultraboosted particle, then we may replace $\Phi$ in \eqref{eqn:BrinkmannGauge} with~\cite{Jackiw:1991ck} 
\begin{align}\label{eqn:ParticleSource}
    \Phi(x^{\lcperp})
    =
    \frac{Q}{4\pi} \log(\mu^{2} r^{2})
    \,,
\end{align}
in which $\mu$ is an arbitrary  scale which yields only pure gauge terms and drops out of physical observables. 
This is the `shockwave' potential which, as discussed in the introduction, arises in high-energy scattering and in gravity. 
Though the potential coincides, naturally, with that of a charged, massless particle~\cite{Bonnor:1969rb}, an example physical situation in which \eqref{eqn:ParticleSource} would be useful is in describing the field of a heavily boosted ion of charge $Q$, which, due to its large mass, is essentially undeflected in collision with an electron. 
Similarly, in employing \eqref{eqn:BrinkmannGauge} as a fixed background, we are assuming that the ultraboosted beam is undisturbed by its interaction with the probe. 
This would seem to be in-line with the desire for small disruption parameter in beam-beam collisions for studying strong-field QED~\cite{Yakimenko:2018kih}; we note also that~\cite{Yakimenko:2018kih} uses electron beams boosted to 125~GeV, that is 
$99.9999999992\%$ the speed of light, suggesting that the ultra-boost should be a good approximation.

In what follows we leave $\Phi$ unspecified as far as possible, so that our results hold for beams, particles, or other sources which yield a potential of the form $A_\mu$ in \eqref{eqn:BrinkmannGauge}.
For example, the case $\Phi \sim r^2$ for \emph{all} $r$ is studied in~\cite{Hollowood:2015elj} (where, note, it is referred to as the `beam' shockwave) in the context of causality violation and UV completion.

\subsection{Classical particle dynamics in ultraboosted Coulomb fields~\label{sec:Lorentz}}

It is useful for what follows to consider the classical motion of a particle, of charge $e$ and mass $m$, scattering on the ultrarelativistic beam above. 
We must solve the Lorentz force equation
\begin{align}\label{eqn:Lorentz}
    m \ddot{x}_{\mu}
    =
    e
    F_{\mu\nu}
    \dot{x}^{\nu}
    \,,
\end{align}
for the orbit $x^{\mu}$ and kinematic momentum $\pi_{\mu} = m \dot{x}_{\mu}$, where the field strength is \eqref{eqn:FieldStrength} and a dot represents differentiation with respect to proper time $\tau$. 
It is clear from \eqref{eqn:FieldStrength} that the only interaction is at the instant $x^\LCm = 0$, when the particle encounters the field. The strength of this interaction is dependent on the location of the particle in the transverse plane, i.e.~on its impact parameters relative to the beam, due to the dependence of $F_{\mu\nu}$ on $x^\LCperp$. 
Before and after the instant $x^\LCm=0$, though, $F_{\mu\nu}=0$ and the motion is free. 

Contracting \eqref{eqn:Lorentz} with $n^{\mu}$ we find immediately that $n \cdot \ddot{x} =0$, implying that the momentum component $n\cdot \pi$ is conserved and equal to its initial value -- we write $p_\mu$ for the initial momentum. 
Integrating up, $n \cdot {x} = n\cdot p\, \tau/m$, so the orbit can be parametrised by lightfront time $x^\LCm$. 
Turning to the transverse coordinates, the momentum obeys
\begin{align}\label{eqn:TransverseMomentumEOM}
    \partial_{\lcm}
    \pi_{\lcperp}
    =
    -
    e\,
    \delta(x^{\lcm})
    \partial_{\lcperp}
    \Phi(x^{\lcperp})
    \,,
\end{align}
and hence the particle momentum is kicked as it crosses the plane at $x^\LCm=0$, with the strength of this kick dependent on \textit{where} in the transverse plane the particle is when it crosses; let this position be~$b^\LCperp$. 
Then to solve \eqref{eqn:TransverseMomentumEOM} one simply patches across the discontinuity at $x^\LCm=0$, 
\begin{align}\label{eqn:TransverseMomentum}
    \pi_{\lcperp}(x^\LCm)
    = 
    \left.\begin{cases}
            p_{\lcperp} 
            - 
            e {\partial}_{\lcperp} \Phi(b^{\lcperp})  & x^\lcm > 0 \\
            p_{\lcperp} & x^{\lcm} < 0
        \end{cases}
    \right\}
    = 
    p_{\lcperp} 
    - 
    e \theta(x^{\lcm}) {\partial}_{\lcperp} \Phi(b^{\lcperp}) \;.
\end{align}
Integrating up once more yields the transverse orbit,
\begin{align}
    x^{\lcperp} (x^\LCm)
    = 
    \int\!\ud \tau 
    \frac{1}{m} \pi^{\lcperp} 
    = 
    \int\!\ud x^{\lcm} \frac{1}{n\cdot p} \pi^{\lcperp} 
    = 
    b^{\lcperp} 
    +
    \frac{p^{\lcperp}}{p^{\lcp}} x^{\lcm} 
    - 
    e\, x^{\lcm} \theta(x^{\lcm}) {\partial}^{\lcperp} \Phi(b^{\lcperp}) \;,
\end{align}
where the constants of integration are fixed by consistency with \eqref{eqn:TransverseMomentum}, hence $x^\LCperp(0)~=~b^\LCperp$. 
Note that the transverse position is continuous across the shock.  
The final momentum component $\pi_\LCm$ is determined by the mass-shell condition. 
If we define
\be\label{eqn:RosenPotential}
a_\mu(x) = e\,\theta(x^\LCm)\delta_\mu^\LCperp \partial_\LCperp \Phi(x) \;,
\ee
then the full four-momentum, both before and after interaction with the beam, is conveniently expressed in terms of (abusing notation) $a_\mu(b) \equiv a_\mu(x^\LCm,b^\LCperp)$: 
\begin{align}
    \label{eqn:pi-in-shock}
    \pi_{\mu} 
    = 
    p_{\mu} 
    - 
    a_{\mu}(b) 
    + 
    n_{\mu} 
    \frac{2 a(b)\cdot p - a(b) \cdot a(b)}{2 n\cdot p} 
    \;.
\end{align}         
In summary, a particle experiences a momentum kick as the infinitely boosted beam passes by. 
The strength and direction of the kick is determined entirely by the transverse position of the particle relative to the beam, i.e.~the impact parameter~$b$. 
The orbit of the particle is continuous, and comprises two straight lines (free motion) patched at the moment of interaction. 
As such the orbit exhibits a velocity memory effect~\cite{Dinu:2012tj,Bieri:2013hqa,Zhang:2017geq,Zhang:2017rno}.

\medskip

Let us compare these results with a particle scattering from an ultrashort, `impulsive' plane wave, which, unlike \eqref{eqn:FieldStrength} is a solution of Maxwell's equations (i.e. source-free). 
The potential may be written (including a factor of the coupling for later convenience)
\be\label{eqn:PWBrink}
eA^\text{p.w.}_\mu(x) = -n_\mu \delta(n\cdot x) c_\LCperp x^\LCperp \;,
\ee
in which the two-component vector $c_\LCperp$ encodes the strength and polarisation of the impulse. 
The momentum of a particle crossing the impulse is functionally identical to \eqref{eqn:pi-in-shock}, but with $a_\mu(b) \to \theta(x^\LCm) c_\mu$: the momentum is kicked (and the orbit is continuous), but the kick and memory effect are now `global', i.e.~\textit{independent} of the transverse position of the particle. This is simply because the plane wave field strength is independent of $x^\LCperp$.
The similarities between the shockwave and impulsive plane wave will be useful in the QED calculations below.

\section{Fermion wavefunctions on ultraboosted beam backgrounds \label{sec:Quantum}}

In order to compute amplitudes in a strong background, one requires explicit wavefunctions to represent the on-shell external particles in the scattering process. 
For electrons and positrons, these wavefunctions are determined by solving the Dirac equation coupled to the fixed, classical background. 
For generic backgrounds finding these solutions may be difficult or impossible, but highly symmetric backgrounds often enable exact wavefunctions to be written down~\cite{Bagrov:1990xp}. 
This is the case, for example, for plane waves~\cite{volkov35,DiPiazza:2011tq,Seipt:2017ckc,Heinzl:2017zsr}.

In this section, we show that the ultraboosted beam and shockwave backgrounds introduced in Sec.~\ref{sec:Classical} also allow the determination of exact wavefunctions. 
For ultraboosted beams of the form~\eqref{eqn:BrinkmannGauge} these wavefunctions do not appear to have been systematically studied before. 
While the solution of the scalar wave equations in electromagnetic (and gravitational) shockwaves \eqref{eqn:ParticleSource} is well-covered in the literature (cf., \cite{Dray:1984ha,Jackiw:1991ck,Kabat:1992tb,Lodone:2009qe}), the Dirac equation is less commonly studied (though see~\cite{Meggiolaro:1995cu} for the eikonal calculation). 
As these calculations are perhaps less familiar in the laser physics community -- and as the continuity conditions required are not entirely trivial -- we also cover the shockwave calculation in some detail here.

\subsection{Solving the Dirac equation \label{sec:Dirac}}

We consider the Dirac equation
\begin{align}\label{eqn:Dirac}
    \big(
        i
        \slashed{D}
        -
        m
    \big)
    \psi
    =
    0
    \,,
    \qquad 
    D_{\mu} = \partial_{\mu} +i e A_{\mu} \;,
\end{align}
for fermions in the ultraboosted beam background \eqref{eqn:BrinkmannGauge}, with $\Phi(x^\LCperp)$ left unspecified. 
As this gauge potential is singular, ensuring that the Dirac equation is satisfied on the plane of interaction at   $x^\LCm=0$ is subtle.
It is simpler to first solve \eqref{eqn:Dirac} in a non-singular gauge before transforming back to the form \eqref{eqn:BrinkmannGauge}. 
The reason for returning to the singular gauge at the end of the calculation is that it provides the simplest form of the wavefunctions with which to calculate scattering amplitudes.

To this end, observe that $eA_\mu(x)=a_\mu(x)=e\theta(x^{\LCm})\,\delta^{\LCperp}_{\mu}\partial_{\LCperp}\Phi(x^\LCperp)$ as in \eqref{eqn:RosenPotential} is a valid gauge potential for the field \eqref{eqn:FieldStrength}. 
It is easy to see that this potential is related to \eqref{eqn:BrinkmannGauge} via a gauge transformation generated by $-\theta(x^-)\Phi(x^\LCperp)$. 
In this new gauge, there is no singularity in the potential and, given its form, it is natural to decompose solutions of the Dirac equation as 
\begin{align}\label{eqn:WavefunctionGeneral}
    \psi(x)
    =
    \theta(- x^{\lcm})
    \psi_{<}(x)
    +
    \theta(x^{\lcm})
    \psi_{>}(x)
    \,,
\end{align}
where $\psi_{<}$ and $\psi_{>}$ are, respectively, the wavefunctions above and below the plane of interaction with the field. 
We start by focusing on \emph{incoming} electrons: those $\psi$ which reduce to free electron wavefunctions in the infinite past (all other solutions are given below). 
As the potential $a_\mu$ vanishes for $x^\LCm<0$, $\psi_<$ is determined there by its initial condition and hence $\psi_{<}(x) = e^{- i p \cdot x} u_{p}^{s}$ for initial momentum $p_\mu$ and spin $s$. 
Above the plane of interaction, the Dirac equation is solved by any linear combination of functions of the form $e^{- i q \cdot x - i e\Phi(x^{\lcperp})} u_{q}^{r}$, for arbitrary (on-shell) $q_\mu$ and spin $r$: 
\begin{align}\label{eqn:AboveShock}
    \psi_{>}(x)
    =
    \int \frac{\ud q_{\lcp}\ud^{2} q_{\lcperp}}{(2 \pi)^{3}} \Lambda^{r}(q)
    e^{- i q \cdot x - i e\Phi(x^{\lcperp})}
    u_{q}^{r}
    \,,
\end{align}
where the $\Lambda^{r}(q)$ are unknown coefficients. 
These are determined by ensuring that the Dirac equation is satisfied \textit{everywhere}, in particular at the $x^\LCm=0$ lightfront itself.  
With $\psi_<$ below and $\psi_>$ above, satisfying the Dirac equation implies a continuity condition at $x^\LCm=0$:
\begin{align}\label{eqn:DiracZero}
    \big(
        i
        \slashed{D} - m
    \big) \psi
    =
    -
    \delta(x^{\lcm})
    \slashed{n} \,
    \psi_{<}(x)
    +
    \delta(x^{\lcm})
    \slashed{n}\,
    \psi_{>}(x)
    \overset{!}{=}
    0
    \,.
\end{align}
This may be written as
\begin{align}\label{eqn:Matching}
    e^{- i p_\LCp x^\LCp- i p_\LCperp x^\LCperp}
    \slashed{n}
    u^{s}_{p}
    =
    \int \frac{\ud q_\LCp \ud^{2} q_{\lcperp}}{(2 \pi)^{3}}
    e^{- i q_\LCp x^\LCp- i q_\LCp x^\LCp}
    e^{- i e\Phi(x^{\lcperp})}
    \Lambda^{r}(q)
    \slashed{n}
    u_{q}^{r}
    \,.
\end{align}
Rearranging and taking a Fourier transform yields
\begin{align}
    \label{eqn:Coefficients}
    \Lambda^{r}(q)
    \slashed{n}
    u_{q}^{r}
            &=
            (2 \pi)
            \delta(p_{\lcp} - q_{\lcp})
            W(p-q)
            \slashed{n}
            u_{p}^{s} \;,
\end{align}
in which we have defined the weight
\begin{align}\label{eqn:Kernel}
    W(q)
    \equiv
    \int \ud^{2} y^{\lcperp}
    e^{- i q_{\lcperp} \cdot y^{\lcperp}}
    e^{i e\Phi(y^{\lcperp})}
    \,.
\end{align}
Note that a naive imposition of continuity of $\psi(x)$ at $x^\LCm=0$ would have led to \eqref{eqn:Coefficients} \textit{without} the factor of $\slashed{n}$: those equations have no solution. 
The factor of  $\slashed{n}$ reduces the effective degrees of freedom in the spinors, though, and allows for a solution. 
To see how, note the useful result that for an on-shell $q_\mu$ with $q_\LCp = p_\LCp$, we have
\be\label{eqn:VolkShift}
u_q = \bigg(1 + \frac{\slashed{n}(\slashed{p}-\slashed{q})}{2n\cdot p}\bigg)u_{p}\,.
\ee
Multiplying \eqref{eqn:Coefficients} with ${\bar u}_q$, then using \eqref{eqn:VolkShift} and the Gordon identity, we finally obtain
\begin{align}\label{eqn:CoefficientsDelta}
    \Lambda^{r}(q)
    =
    (2 \pi)
    \delta(p_{\lcp} - q_{\lcp})
    W(p - q)
    \delta^{r s}
    \,,
\end{align}
thereby completing the solution of the Dirac equation. 
Finally, we transform back to the gauge \eqref{eqn:BrinkmannGauge}: this is achieved by multiplying $\psi$ by the phase $e^{i e \theta(x^\LCm) \Phi} $ which simply removes the momentum independent phase $e^{- i e\Phi}$ from $\psi_>$ in \cref{eqn:AboveShock}. 
Thus, the wavefunction
\begin{align}\label{eqn:ElectronIn}
    \psi_{p, \text{in}}(x)
    =
    \theta(- x^{\lcm})
    u_{p}
    e^{- i p \cdot x}
    +
    \theta(x^{\lcm})
    \int \frac{\ud^{2}q_{\lcperp}}{(2 \pi)^{2}}
    W(p - q)
    e^{- i q \cdot x}
    u_{q}\Big|_{q_\LCp = p_\LCp}
    \,,
\end{align}
will be used for incoming electrons in all subsequent scattering calculations. 
The wavefunction \eqref{eqn:ElectronIn} is the analogue of the Volkov wavefunction for a particle interacting with a plane wave background~\cite{volkov35}. 
Its properties are as follows.

After interacting with the field, the wavefunction becomes a superposition of \textit{free} single electron wavefunctions; this is because the potential \eqref{eqn:BrinkmannGauge} vanishes everywhere except at $x^\LCm=0$. 
We have a single-particle wavefunction because (as for plane waves) there is no spontaneous (Schwinger) pair production in ultraboosted beams: $F^{\mu\nu}F_{\mu\nu} = F^{\mu\nu}{\tilde F}_{\mu\nu}=0$ as is easily seen from \eqref{eqn:FieldStrength}. 
We have conservation of momentum $p_\LCp$, as in the classical theory (seen already through the delta function in \eqref{eqn:Coefficients}). 
The superposition sums over transverse momenta which means, like the classical theory, that the particle momentum is changed as it crosses the field; however, unlike the classical particle it may be kicked to \emph{any} momentum $q_\mu$. 
The probability amplitude for this transition is, as we will make clear below, just $W(p-q)$. 
The spin state of the particle is preserved across the shock, hence we have dropped explicit spin labels.

For completeness and later use, we list the remaining incoming and outgoing states:
\begin{align}\label{eqn:ElectronOut}
    e^\LCm\,\,\text{ out:}\qquad
    \bar{\psi}_{p, \text{out}}(x)
    =
    \theta(x^{\lcm})
    \bar{u}_{p}\,
    e^{i p \cdot x}
    +
    \theta(- x^{\lcm})
    \int \frac{\ud^{2}q_{\lcperp}}{(2 \pi)^{2}}
    W(q - p)
    e^{i q \cdot x}
    \bar{u}_{q}
    \;,
\end{align} 

\begin{align}\label{eqn:PositronIn}
    e^\LCp\,\,\text{ in:}\qquad
    \bar{\varPsi}_{p, \text{in}}
    =
    \theta(-x^{\lcm})
    \bar{v}_{p}\,
    e^{-i p \cdot x}
    +
    \theta(x^{\lcm})
    \int \frac{\ud^{2}q_{\lcperp}}{(2 \pi)^{2}}
    W^{\dagger}(q - p)
    e^{- i q \cdot x}
    \bar{v}_{q}
    \;,
\end{align}
\begin{align}\label{eqn:PositronOut}
    e^\LCp\,\,\text{ out:}\qquad
    \varPsi_{p, \text{out}}
    =
    \theta(x^{\lcm})
    v_{p}\,
    e^{i p \cdot x}
    +
    \theta(- x^{\lcm})
    \int \frac{\ud^{2}q_{\lcperp}}{(2 \pi)^{2}}
    W^{\dagger}(p - q)
    e^{i q \cdot x}
    v_{q}
    \;,
\end{align}
in which $q_\LCp = p_\LCp$ throughout. 
The outgoing states are definite momentum eigenstates \textit{above} the field, and superpositions below it -- this means that the field can kick a range of different momentum states into the definite asymptotic state. 
The argument of $W$ is flipped in outgoing states relative to incoming. 
Positrons come with $W^\dagger$, rather than\footnote{As a consistency check, note that to go from electrons to positrons (\textit{both} incoming or both outgoing) we change  $u \to {\bar v}$ or ${\bar u} \to v$ as in vacuum, and flip the sign of $e$, which is equivalent to replacing $W(p) \to W^\dagger(-p)$.} $W$. 

These solutions hold for arbitrary `profile' functions $\Phi(x^\LCperp)$, and their associated electromagnetic fields. 
In particular, the solutions are exact in the \textit{strength} of these fields as defined by the total charge of the bunch in \eqref{eqn:BrinkmannGauge} or of the single ultraboosted point charge in \eqref{eqn:ParticleSource}.

\subsection{Elastic scattering\label{sec:2Point}}

As a first example of a scattering calculation, which will shed light on the physical interpretation of $W$, consider the $1 \to 1$ scattering (without emission) of an electron off an ultra-relativistic beam background \eqref{eqn:BrinkmannGauge}. 
Suppose the electron has initial momentum $p_\mu$ and scatters off the beam to final momentum $p'_\mu$. 
The probability amplitude for the transition is just the overlap, in the asymptotic future, of $\psi_{p,\text{in}}$ with a free electron state of momentum $p'_\mu$,
\begin{align}\label{eqn:2pointnoncovariant}
    \lim_{x^{\lcm}\to\infty} \int\!\ud^2 x^{\lcperp}\ud x^{\lcp} \, \bar{u}_{p'} e^{ip'\cdot x} \gamma^{\lcm} \psi_{p,\text{in}}(x) = \bar{u}_{p'} \slashed{n} u_{p}\,2\pi\,{\delta(p_\LCp-p'_\LCp)}\,  W(p - p') \,,
\end{align}
and is equal to $W$ decorated with a spin factor.  
More formally, the $S$-matrix element for $1 \to 1$ scattering is obtained from applying LSZ reduction to both ends of the background-dressed propagator; applying LSZ first to the `incoming' end of the propagator will yield
\begin{align}\label{eqn:2point}
    S_{fi} 
    = 
    -
    i 
    \int\!\ud^4 x \, 
    \bar{u}_{p'} 
    e^{ i p' \cdot x} 
    \big(i \slashed{\partial} - m \big) 
    \psi_{p,\text{in}}(x)
    \;,
\end{align}
with the remaining structure being amputation for the outgoing leg. 
Evaluating the derivatives, and using the explicit form of the wavefunction \eqref{eqn:ElectronIn} we find
\begin{align}\label{eqn:2pointexpanded}
    S_{fi} 
    &=
    \bar{u}_{p'} \slashed{n}u_{p}\, 2\pi\,\delta(p_\LCp-p'_\LCp)\, W(p-p') 
    -
    \bar{u}_{p'} \slashed{n}u_{p}\, (2\pi)^3\, \delta^{3}_{\lcp,\lcperp}(p-p') 
    \\
    &
    \label{eqn:impact1}
    =     \bar{u}_{p'} 
    \slashed{n}
    u_{p}\, 
    2\pi\, 
    \delta(p_\LCp-p'_\LCp) 
    \int \!\ud^2 b_{\lcperp} 
    e^{i b^{\lcperp} \cdot (p'_{\lcperp} - p_{\lcperp})}
    \big(e^{i e\Phi(b^\LCperp)}-1 \big)\;,
\end{align}
where $\delta^3_{\lcp,\lcperp}(p)$ denotes three delta functions in the $p_+$, $p_\lcperp$ components. 
This differs from \eqref{eqn:2pointnoncovariant} in the second term, which is a subtraction of the forward-scattering contribution in which the particle and beam do not interact at all. 
This is made explicit in \eqref{eqn:impact1} using the `impact parameter' representation\footnote{There is no contradiction between \eqref{eqn:2pointnoncovariant} and \eqref{eqn:2pointexpanded}. The former is the textbook starting point for LSZ which \textit{includes} the forward scattering contribution. This is subtracted when going to the covariant `LSZ proper' expressions, \eqref{eqn:2point}. The subtraction is a well-known part of the eikonal approach~\cite{Lodone:2009qe} but it is sometimes thought to be missed in the background field approach -- we see that it is in fact included.}.

Squaring up and summing/averaging over final/initial spins, the differential cross-section (away from forward scattering) may be expressed in terms of the momentum transfer $q_\LCperp := p_\LCperp - p'_\LCperp$ and lightfront momentum fraction $z:=p'_\LCp/p_\LCp$ of the outgoing electron as
\be\label{X1-tripple}
	\frac{\ud^2 \sigma}{\ud^2 q_\LCperp\ud z}  = \delta(z-1)\frac{|W(q)|^2 }{(2\pi)^2} \;.
\ee
As $W$ will continue to play a key role in the processes to be considered later (nonlinear Compton and nonlinear Breit-Wheeler) we discuss its properties in two specific cases. 

Consider first a single ultra-boosted particle of charge~$Q$, that is we take~$\Phi(x^\LCperp)$ as in \eqref{eqn:ParticleSource}, describing a shockwave. 
Then the integrals in \eqref{eqn:Kernel} can be performed and the resulting expression, call it $W_1(q)$, is well-known in the literature~\cite{tHooft:1987vrq,Jackiw:1991ck,Giudice:2001ce}; we write it here in a slightly more revealing form. 
Defining $\xi:=eQ/(4\pi)$, then
\begin{align}\label{eqn:KernelParticleIntegrated}
    W_\text{1}(q) 
    =
    \frac{4\pi \xi}{|q_\LCperp|^2} \times i \bigg(\frac{4\mu^2}{|q_\LCperp|^2}\bigg)^{2i \xi} \frac{\Gamma(i \xi)}{\Gamma(-i \xi)} = |W(q)| \times e^{i\phi} \;.
\end{align}%
The first factor is the modulus of $W$, while everything after the `$\times$' is a pure phase, as illustrated by the second equality. 
From this it is evident that $\mu$ will drop out of probabilities, cross-sections, etc, upon taking the modulus of $W_1$ squared.  
Dimensionless $\xi$ characterises the strength of the interaction between the probe and the background. 
While \eqref{eqn:KernelParticleIntegrated} clearly contains terms of all orders in $\xi$, the cross-section for elastic scattering is simply, from (\ref{X1-tripple}),
\be\label{eqn:2PointProb}
	\frac{\ud^2 \sigma}{\ud^2 q_\LCperp}  = \frac{4\xi^2 }{|q_\LCperp|^4} \;,
\ee
for \textit{any value} of $\xi$, as is well known~\cite{tHooft:1987vrq}.

Now consider the scattering of an electron from the ultraboosted beam~\eqref{eqn:BrinkmannGauge}, continuing to write $\xi=eQ/(4\pi)$ but where $Q$ now represents the total charge of the beam. 
Recall that $r_0$ is the radius of the beam. 
Clearly $W$ will differ from the single-charge result $W_1$ only at high momentum transfer, since  the potential \eqref{eqn:BrinkmannGauge} differs from \cref{eqn:ParticleSource} only in its short-distance structure.  
We may write
\begin{align}\label{eqn:KernelBeamIntegrated}
    W(q)
    =
    e^{i \xi}
    W_{1}(q)
    +
    W_{2}(q)
    \,,
\end{align}
in which $W_1$ is the shockwave result \eqref{eqn:KernelParticleIntegrated} with $\mu = 1/r_0$, and
\begin{align}\label{eqn:W2}
    W_{2}(q)
    :=
    2 \pi r_{0}^{2}
    \int_{0}^{1} \ud x \, x \,
    J_{0}(|q_\LCperp| r_{0} x)
    \Big(
        e^{i \xi x^{2}}
        -
        e^{i \xi + i \xi \log x^{2}}
    \Big)
    \,.
\end{align}
Of the two terms in $W_2$, the second can be expressed in terms of the hypergeometric function ${}_{1}F_{2}$, but (to the best of our knowledge) there is no exact result for the first integral. 
We can easily analyse its particular behaviour in various limits, however. 
Of particular interest is the high-field, $\xi \gg 1$, behaviour.

\begin{figure}[t!]
    \includegraphics[width=0.45\textwidth]{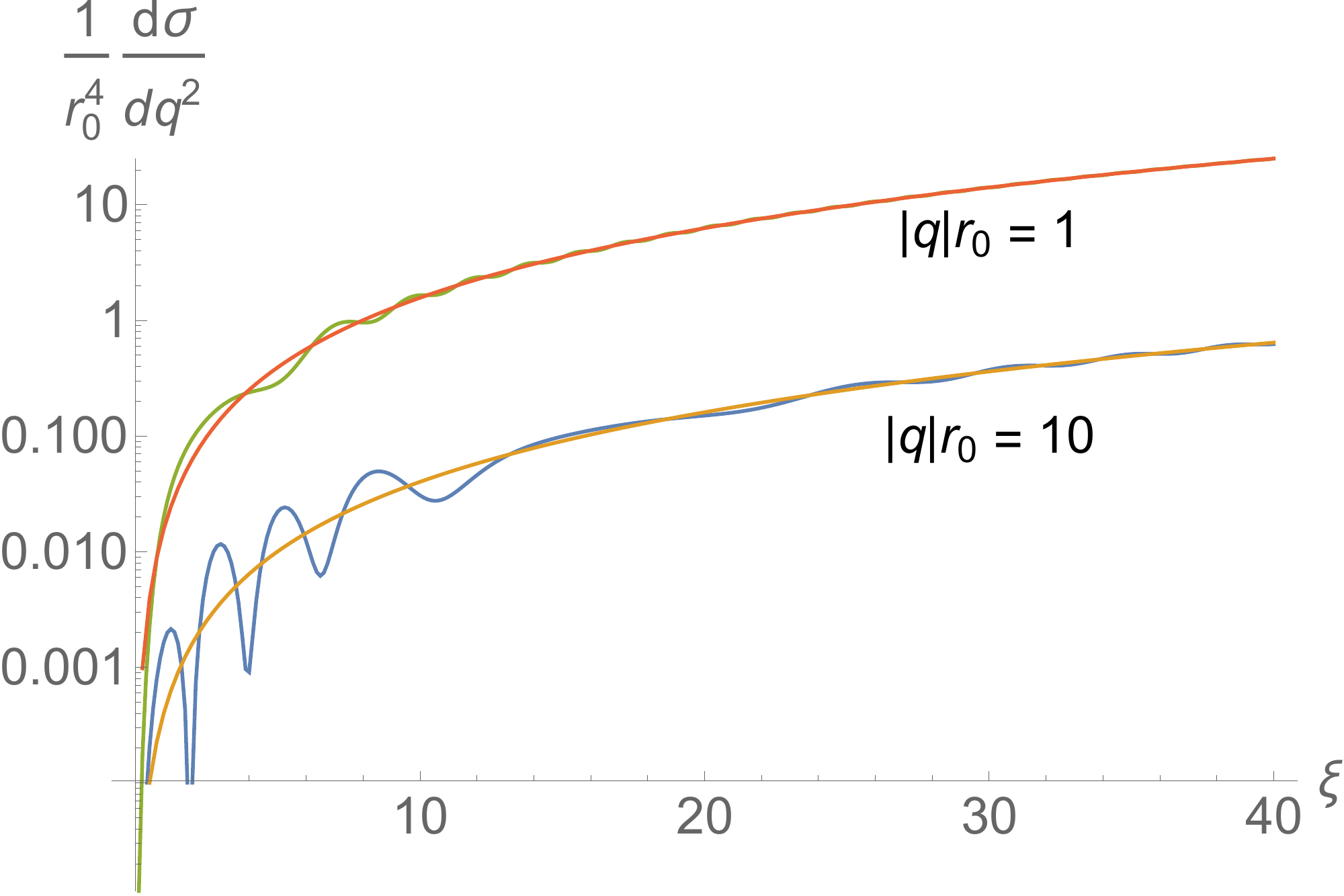}
    \qquad
    \includegraphics[width=0.45\textwidth]{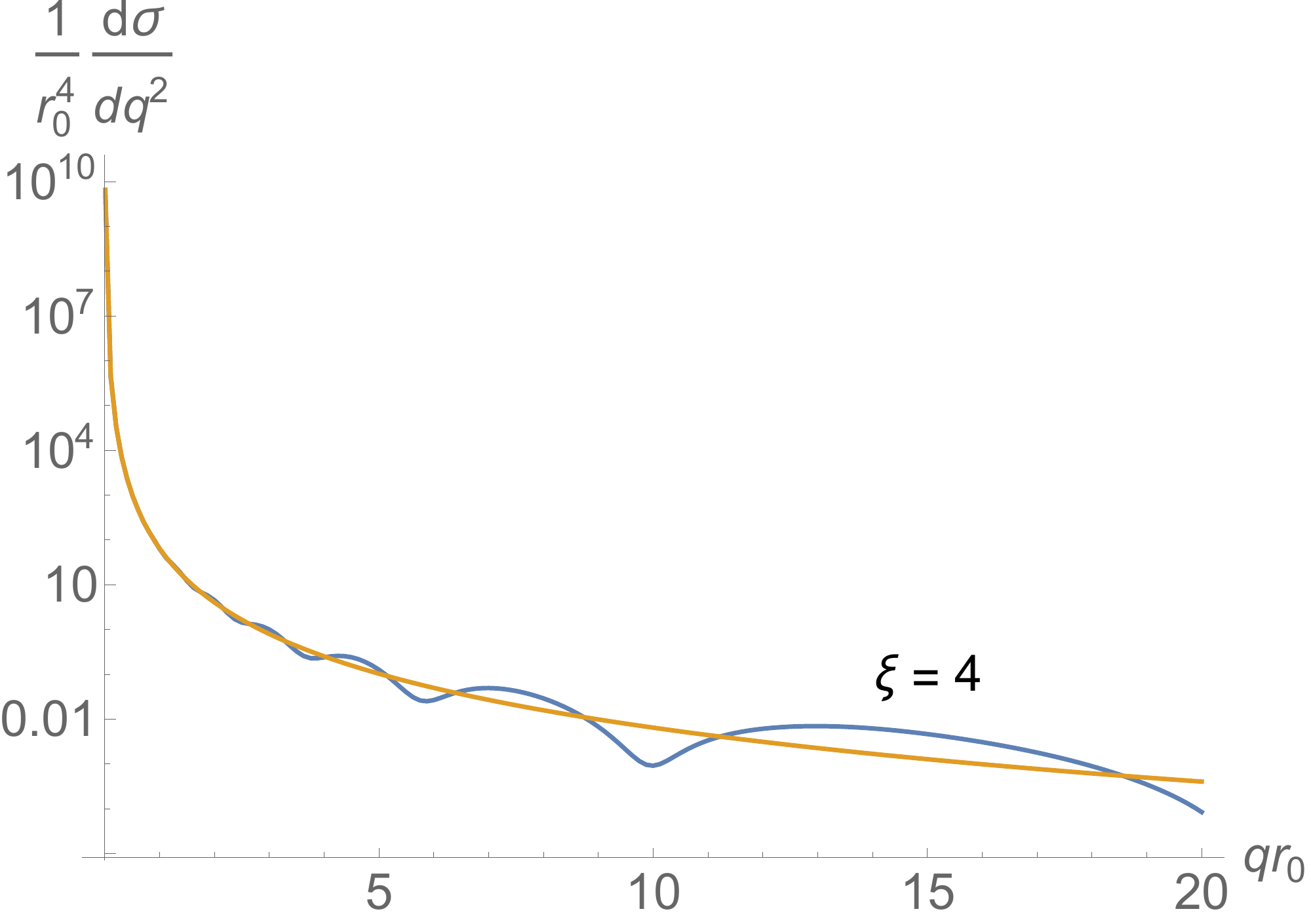}
    \caption{\label{FIG:XSEKT} A comparison of the differential cross-section (\ref{X1-tripple}) (integrated over $z$) in the beam and shockwave backgrounds. \textit{Left}: as the total charge of the beam increases, it becomes well-approximated by the shockwave. \textit{Right}: at fixed field strengths, the beam and shockwave cross-sections differ at large momentum transfer $q$, which probes the short-distance structure of the beam.}
\end{figure}

As $\xi$ appears in the exponentials of both terms in $W_{2}$, it suggests that the large $\xi$ limit leads to cancellations through rapid oscillations.  
To analyse this more carefully, we perform some simple manipulations of the integrals.  
In the first term of \eqref{eqn:W2} we use integration by parts to obtain
\begin{align}\label{eqn:W2a}
    \int_{0}^{1} \ud x \, x \,
    J_{0}(|q_\LCperp| r_{0} x)
    e^{i \xi x^{2}}
    =
    \frac{1}{2i \xi}
    \bigg[
        J_{0}(|q_\LCperp| r_{0})
        e^{i \xi}
        -
        1
        +
        |q_{\LCperp}| r_{0}
        \int_{0}^{1} \ud x \,
        e^{i \xi x^{2}}
        J_{1}(|q_\LCperp| r_{0} x)
    \bigg]
    \,.
\end{align}
The terms in square brackets are all bounded in modulus, and this bound is $\xi$-independent. 
Hence (\ref{eqn:W2a}) goes like $1/\xi$ for large $\xi$. 
Doing the same for the second term in (\ref{eqn:W2}) yields
\begin{align}\label{eqn:W2b}
    \frac{e^{i \xi}}{2(1 + i \xi)}
    \bigg[
        J_{0}(|q_\LCperp| r_{0})
        +
        |q_{\LCperp}| r_{0}
        \int_0^1\! \ud x \,
        x^{2+2i \xi}  J_{1}(|q_\LCperp| r_{0} x)
    \bigg] \;,
\end{align} 
in which the terms in square brackets are again (noting the integral limits) bounded in modulus, and we conclude that $W_2(q)$ in (\ref{eqn:W2}) goes at best as $1/\xi$ for $\xi \gg 1$. 
As such, the leading contribution to $W(q)$ in particle-beam scattering at $\xi\gg 1$ comes from the single particle part $W_1$, going like $\xi^2$. 
This is confirmed in Fig.~\ref{FIG:XSEKT}.

\section{Particle-beam scattering from SFQED~\label{sec:3Point}}

With these ingredients, there is nothing to stop us from proceeding to compute higher-point amplitudes. 
However, before doing so we first find an alternative representation of the fermion wavefunctions which yields an intriguing physical interpretation and simplifies the calculations to be performed.

\subsection{Relation to the Volkov wavefunctions and physical interpretation 
\label{sec:VolkovBasis}}

Recall from section~\ref{sec:Classical} that the physical impact of both an ultraboosted beam and an impulsive plane wave on a particle is to suddenly change its momentum. 
This change is characterised, in both cases, by \textit{two} transverse degrees of freedom. 
For the impulse, these are `global': they are defined by the impulse itself. 
For the beam, though, they are defined by the two transverse impact parameters of the interaction geometry.  
In what follows, we uncover an elegant relation between these two at-first-sight different physical scenarios, in the quantum theory.

The \textit{Volkov wavefunction}~\cite{volkov35} describing an electron with initial momentum $p_\mu$ crossing the impulsive plane wave \eqref{eqn:PWBrink} is~\cite{Ilderton:2019vot,Ilderton:2019ceq}
\begin{align}\label{eqn:VolkovImpulse2}
    \psi_{p,\text{in}}(x) 
    &
    =
    \theta(-x^{\lcm}) u_{p} e^{-ip\cdot x} 
    + 
    \theta(x^{\lcm})
    \,
    \phi_p(x;c)
    \,,
\end{align}
in which\footnote{In the strong-field QED literature, the gauge \eqref{eqn:RosenPotential} is almost universally used for the Volkov wavefunctions, as explained in~\cite{Dinu:2012tj}. The difference between that and \eqref{eqn:VolkovImpulse} is simply the leading exponential factor in the latter.} 
\begin{align}\label{eqn:VolkovImpulse}
    \phi_p(x;c)
    &
    =
    \, 
    e^{i c\cdot  x}\,
    e^{-i p\cdot x - i \frac{2c\cdot p- c^2}{2n\cdot p} x^\LCm} \bigg(1 + \frac{\slashed{n}\slashed{c}}{2n\cdot p}\bigg)u_{p}
    \,.
\end{align}
Now we take the incoming electron wavefunction $\psi_{p, \text{in}}$ in the beam from \eqref{eqn:ElectronIn} and make the change of variables $q_{\lcperp} \to c_{\lcperp} = p_{\lcperp}  - q_{\lcperp}$. 
Using the relation \eqref{eqn:VolkShift} we immediately find
\begin{align}\label{eqn:VolkovElectronIn}
    \psi_{p,\text{in}}(x) 
    &
    =
    \theta(-x^{\lcm}) u_{p} e^{-ip\cdot x} 
    + 
    \theta(x^{\lcm}) \int\!\frac{\ud^2 c_{\lcperp}}{(2\pi)^2} \, 
    W(c) 
    \,
    \phi_p(x;c)
    \,,
\end{align}
Hence, after crossing the beam, the electron becomes a superposition of Volkov wavefunctions, summed over with weight $W(c)$. 

On the one hand, this is not surprising, since the Volkov wavefunctions are a complete set. 
On the other hand, it provides an alternative interpretation of physics in the beam background: every time the electron crosses the $x^\LCm=0$ lightfront, it sees a different plane wave, defined by parameters $c_\LCperp$, with $W(c)$ essentially being the amplitude for which impulsive wave is seen. 
(This interpretation is consistent with the fact that spin is also unchanged when crossing a plane wave~\cite{Ilderton:2020gno}.) 
Recalling that the beam is sourced, it is interesting to note that \eqref{eqn:VolkovElectronIn} also expresses an \textit{all-loop} amplitude in terms of all-orders \textit{tree level} amplitudes, since an external plane wave is a coherent state of free photons~\cite{KibbleShift,Frantz,Gavrilov:1990qa}. 
Finally, for outgoing electrons we may similarly write
\begin{align}\label{eqn:VolkovElectronOut}
    \bar{\psi}_{p,\text{out}}(x) 
    & 
    =
    \theta(x^{\lcm}) \bar{u}_{p} e^{i p \cdot x} 
    + 
    \theta(-x^{\lcm}) \int\!\frac{\ud^2 c_{\lcperp}}{(2\pi)^2} \, 
    W(c) \,
    \overline{{\phi_{p}}}(x;-c)
    \,.
\end{align}

\subsection{Nonlinear Compton scattering \label{sec:NLC}}
%
\begin{figure}[t!]
\includegraphics[width=0.55\textwidth]{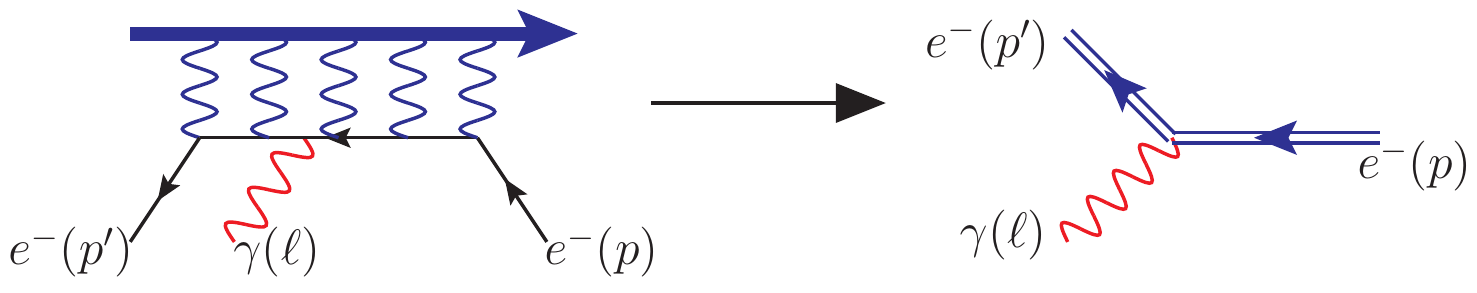}
\caption{\label{fig:ALL-3-PT} \textit{Left}: a probe electron scatters off an ultraboosted beam and emits a photon. \textit{Right}: the fields of the beam are treated as a fixed external field, in which the electron moves; double lines indicate that the interaction of the electron with the background is treated exactly. This is `nonlinear Compton scattering'.}
\end{figure}
We now turn to the investigation of the three-point `nonlinear Compton scattering' amplitude: that is, photon emission from an electron crossing a strong electromagnetic background, here an ultraboosted beam or shockwave. 
Using \eqref{eqn:VolkovElectronIn} we will be able to relate this directly to nonlinear Compton as studied in laser-matter interactions, for reviews see~\cite{DiPiazza:2011tq,Seipt:2017ckc}.

We consider an electron scattering from momentum $p_\mu$ to momentum $q_\mu$ with the emission of a photon of momentum $\ell_\mu$ and polarisation $\epsilon_{\mu}$,
as shown in  \cref{fig:ALL-3-PT}, where double lines represent the incoming and outgoing electron wavefunctions \cref{eqn:ElectronIn,eqn:ElectronOut}. 
The $S$-matrix element is
\begin{align}\label{eqn:NLCAmplitude}
    S^\text{\,beam}_{fi}
    =
    &
    -
    i e
    \int \ud^{4} x \;
    \bar{\psi}_{q,\text{out}}(x)
    {\bar{ \slashed{\epsilon}}}
    e^{i \ell\cdot x}
    \psi_{p,\text{in}}(x)
    \,.
\end{align}
We immediately express this in terms of Volkov wavefunctions using \eqref{eqn:VolkovElectronOut} and \eqref{eqn:VolkovElectronOut} in order to relate the amplitude in the ultraboosted beam to the amplitudes in plane waves, for which the expression is given by \eqref{eqn:NLCAmplitude} but where $\psi$ is now the relevant incoming or outgoing \textit{Volkov} wavefunction, recall \eqref{eqn:VolkovElectronIn}. 
We find 
\begin{align}  
    \label{eqn:shock2pw1}
    S^\text{\,beam}_{fi}
    & =  \int\!\frac{\ud^2 c_\LCperp}{(2\pi)^2} W(c) \,\bigg[-ie\!\int\!\ud^4 x\, \theta(x^\LCm) {\bar u}_q e^{iq\cdot x}\, {\bar{\slashed{\epsilon}}}
        e^{i \ell\cdot x}
        \phi_p(x;c) + \theta(-x^\LCm) \overline{\phi_q}(x;-c) {\bar{\slashed{\epsilon}}}             e^{i \ell\cdot x}
    e^{-ip\cdot x} u_p \bigg]
    \\
    & =
    \int\!\frac{\ud^2 c_\LCperp}{(2\pi)^2} W(c) \, S^{\text{\,p.w.}(c)}_{fi} \;,
\end{align}
so that the scattering amplitude is exactly equal to an average of plane wave amplitudes in the background \eqref{eqn:PWBrink}, weighted with $W(c)$; it is as if the fields of the beam act as a stochastic background, with the scattering event being obtained by averaging over the same event in plane waves. 
Furthermore, it follows from \eqref{eqn:shock2pw1} that one can lift, wholesale, existing amplitudes in impulsive plane wave backgrounds~\cite{Ilderton:2019vot} describing laser-particle interactions and from them obtain amplitudes in ultraboosted beams, describing beam-particle interactions. 
For completeness we nevertheless outline the evaluation of $S_{fi}^\text{\,beam}$.

The coordinate integrals in \eqref{eqn:NLCAmplitude} or \eqref{eqn:shock2pw1} are straightforward. 
The integrals over $(x^{\lcp},x^{\lcperp})$ give momentum conserving $\delta$-functions, as they are the same as for the plane wave case where $p_\LCp$ and $p_\LCperp$ is conserved. 
These $\delta$-functions appear under the $\ud^2c$ integral. 
As indicated, the $S$-matrix element is split into contributions above and below the beam field, and in each part the integral in lightfront time $x^{\lcm}$ is trivial. 
One finds
\begin{align}\label{eqn:NLCAmplitudeIntegrated}
    S^\text{\,beam}_{fi} 
    &= e
    \int \! \frac{\ud^{2} c_{\lcperp}}{(2\pi)^{2}}
    W(c)\, 
    (2\pi)^{3}
    \delta_{\lcp,\lcperp}^{3}(q + \ell + c - p)(p_{\lcp} - \ell_{\lcp})
    \bigg[
        \frac{\bar{u}_{q} {\bar{\slashed{\epsilon}}} V_p u_{p}}{\ell\cdot \pi}
        -
        \frac{\bar{u}_{q} V_q {\bar{\slashed{\epsilon}}} u_{p}}{\ell\cdot p}
    \bigg]
    \,,
\end{align}
in which
\begin{align}\label{eqn:DressedSpin}
    V_{p}
    =
    \Big(
        1
        +
        \frac{\slashed{n} \slashed{c}}{2 n \cdot p}
    \Big) 
    \,, \qquad  \pi^{\mu}
    = p^{\mu} - c^{\mu} + \frac{2 c \cdot p - c^{2}}{2 n \cdot p}
    n^{\mu}
    \,.
\end{align}
While the remaining integrals in \eqref{eqn:NLCAmplitudeIntegrated} can be performed immediately using the $\delta$-functions, we keep them for now, in order to pursue the relation to plane wave quantities. 

A natural observable to calculate from the amplitude above is the (differential) cross-section~\cite{Lodone:2009qe}. 
The natural observable in a plane wave background is, however, the (differential) probability~\cite{Ilderton:2019bop}. 
By `natural' we mean those quantities which can be expressed directly in terms of the scattering amplitudes, without having to explicitly retain particle densities or wavepackets. 
(These differ in between backgrounds due to the different symmetries of those backgrounds). 
Just as the amplitudes in the two backgrounds are related by \eqref{eqn:shock2pw1}, we now show that there is a simple relation between the cross-section in a shockwave and the probability in an impulsive plane wave. 

Including a wavepacket with an explicit impact factor for the incoming electron, we square up, sum/average over final/initial spins, integrate out the final state momenta, and also the impact parameter to obtain the cross-section. 
We find that the cross section $\sigma$ is
\be\label{eqn:sigmaprob2}
    \sigma = \int\!\frac{\ud^2 c_\LCperp}{(2\pi)^2} \, |W(c)|^2\, \mathbb{P}_\text{nlc}(c) \;,
\ee
in which $\mathbb{P}_\text{nlc}(c)$ is the probability of nonlinear Compton scattering in the impulsive plane wave background $eA^\text{p.w.}_\mu(x) = -n_\mu \delta(n\cdot x) c_\LCperp x^\LCperp$ (as in \eqref{eqn:PWBrink}). 
Expressed in terms of the emitted photon momenta  $\ell_\LCperp$ and $s := \ell_{\lcp}/p_{\lcp}$, the probability is~\cite{Ilderton:2019vot}
\begin{align}\label{eqn:NLCImpulseProbability}
    \mathbb{P}_\text{nlc}(c)        
    =
    -
    \frac{\alpha m^{2}}{4 \pi^{2}}
    \int \ud^{2} \ell_{\lcperp}
    \int_{\epsilon_{\text{IR}}}^{1} \frac{\ud s}{s} (1 - s) \;
    \bigg[
        \frac{1}{(\ell\cdot \pi)^{2}}
        +
        \frac{1}{(\ell\cdot p)^{2}}
        -
        \frac{2}{(\ell\cdot \pi) (\ell\cdot p)}
        \Big(
            1
            +
            \frac{g(s)|c_{\lcperp}|^{2}}{m^{2}}
        \Big) 
    \bigg]
    \,,
\end{align}
in which $\epsilon_{\text{IR}}$ is an infrared cutoff and
\begin{align}\label{eqn:gs}
    g(s)
    =
    \frac{1}{2}
    +
    \frac{s^{2}}{4 (1 - s)}
    \,.
\end{align}
Combining \eqref{eqn:NLCImpulseProbability} with \eqref{eqn:sigmaprob2}, we can integrate out the photon variables exactly and so express the differential cross section in terms of the momentum transfer~$c$ from the beam.  
Keeping only the terms which are finite or singular as the IR cutoff is removed, and writing $c_{0} := |c_{\lcperp}|/m$ and $c_\star := \sqrt{c_0^2+4}$ to compactify notation, one obtains
\begin{align}\label{eqn:NLCImpulseProbabilityIntegrated}
    \mathbb{P}_\text{nlc}(c)
    =
    \frac{2 \alpha}{\pi}
    \bigg[
        \big(1 + \log\epsilon_{\text{IR}}\big) 
        -
        \frac{4}{c_{0} c_\star}
        \tanh^{-1}\bigg(
            \frac{c_{0}}{c_\star}
        \bigg) 
        \bigg(
            (1
            +
            \tfrac{1}{2}
            c_{0}^{2})(1+\log\epsilon_{\text{IR}})
            -\frac18
            c_0^2
        \bigg)
    \bigg]
    \,,
\end{align}
for the probability in an impulsive plane wave.

We now turn to the total cross-section, which is dominated by its infra-red behaviour. 
Taking, for simplicity, the case of a shockwave (i.e., the beam sourced by a single ultraboosed particle) background \eqref{eqn:KernelParticleIntegrated}, the cross-section is
\begin{align}\label{eqn:NLCSingleParticle}
    \sigma^{\text{shock}} =
    &
    \int \frac{\ud^{2} c_{\lcperp}}{(2\pi)^{2}}
    |W(c_{\lcperp})|^{2}
    \mathbb{P}_\text{nlc}(c)
    =
    \frac{8 \pi \xi^2}{m^{2}}
    \int_{0}^{\infty}
    \frac{\ud c_0}{c_{0}^{3}}\,
    \mathbb{P}_\text{nlc}(c)
    \,.
\end{align}
The remaining integral in $c_{0}$ can be performed exactly, and again requires an IR cut-off $c_{\text{IR}}$. 
Retaining only terms which are divergent or constant when \textit{either} $c_\text{IR}$ or $\epsilon_\text{IR}$ is taken to zero, we find
\begin{align}\label{eqn:NLCIntegratedProbability}
    \sigma^{\text{shock}} =
    \frac{2 \alpha}{3}
    \frac{\xi^2}{m^{2}}
    \Big[8 \log c_\text{IR} \log \epsilon_\text{IR}+5 \log c_\text{IR}-\frac{26 \log \epsilon_\text{IR}}{3}-\frac{17}{3}
    \Big] 
    \,.
\end{align}
The double-logarithm is typical of infra-red behaviour. 
The first, $\log c_\text{IR}$, cuts of the overall momentum transfer and is perhaps the more familiar; it stems from the $1/c^4$ factor in $|W|^2$ which is due to the long-range Coulomb force of the boosted particle generating the shockwave fields. 
The second, $\log\epsilon_\text{IR}$, corresponds to a cutoff on measurement of the lightfront momentum of the emitted photon, relative to that of the initial electron, $n\cdot \ell/n\cdot p$. 
As such it regulates both soft divergences as $\ell_0\to 0$, and \textit{collinear} divergences as $\ell_\mu \to \ell_\LCm n_\mu$, in which the photon is emitted in the direction of motion of (the plane wave and) the shock~\cite{Edwards:2020npu}.

\subsection{Nonlinear Breit-Wheeler pair production \label{sec:NLBW}}

We turn now to the nonlinear Breit-Wheeler process, that is the production of an electron-positron pair, momenta $p_\mu$ and $q_\mu$, from a photon of momentum $\ell_\mu$ and polarisation $\epsilon_\mu$ in the ultraboosted beam background. 
The $S$-matrix element is
\begin{align}\label{eqn:BWAmplitude}
    S_{fi}
    =
    &
    -
    i e
    \int \ud^{4} x \;
    \bar{\psi}_{p,\text{out}}(x)
    \slashed{\epsilon}
    e^{- i \ell\cdot x}
    \varPsi_{q,\text{out}}(x)
    \,.
\end{align} 
From the form of the wavefunctions \eqref{eqn:ElectronOut} and \eqref{eqn:PositronOut}, it is clear that the $S$-matrix element will be quadratic in $W$ below the shock, and independent of $W$ above the shock. 
This is unlike nonlinear Compton, where the whole amplitude was linear in $W$. 
While this poses no real complication, we can maintain an analogous relationship to \eqref{eqn:shock2pw1} at the amplitude level simply by using a different basis of outgoing states. 
Observe that $W$ obeys
\begin{align}\label{eqn:KernelDelta}
    \int \frac{\ud^{2} q_{\lcperp}}{(2\pi)^{2}}
    W(q - p)
    W^{\dagger}(q - k)
    =
    (2\pi)^{2} 
    \delta^{2}_\LCperp(p - k)
    \,,
\end{align}
and hence that we may write the outgoing electron wavefunction \eqref{eqn:ElectronOut} as
\begin{align}\label{eqn:ElectronOutAlt}
    \bar{\psi}_{p, \text{out}}
    =
    \int \frac{\ud^{2} q_{\lcperp}}{(2\pi)^{2}}
    W(q - p)
    \bigg[
        \theta(- x^{\lcm})
        e^{i q \cdot x}
        \bar{u}_{q}
        +
        \theta(x^{\lcm})
        \int \frac{\ud^{2} k_{\lcperp}}{(2\pi)^{2}}
        W^{\dagger}(q - k)
        \bar{u}_{k}\,
        e^{i k \cdot x}
    \bigg]
    \;.
\end{align}
Assuming that we integrate out the electron variables at the level of the probability or cross section, it does not matter if we calculate the amplitude with $\bar{\psi}_{p,\text{out}}$ or the new wavefunction in the square brackets of \eqref{eqn:ElectronOutAlt}; this is because the integration against $W$ in \eqref{eqn:ElectronOutAlt} is -- using \eqref{eqn:KernelDelta} -- a unitarity transformation, and hence either wavefunction yields the same total probability. 
Further, using the new wavefunction means that the $S$-matrix element is \emph{linear} in $W$, as it was for nonlinear Compton scattering. 

\begin{figure}[t!]
    \includegraphics[width=0.7\textwidth]{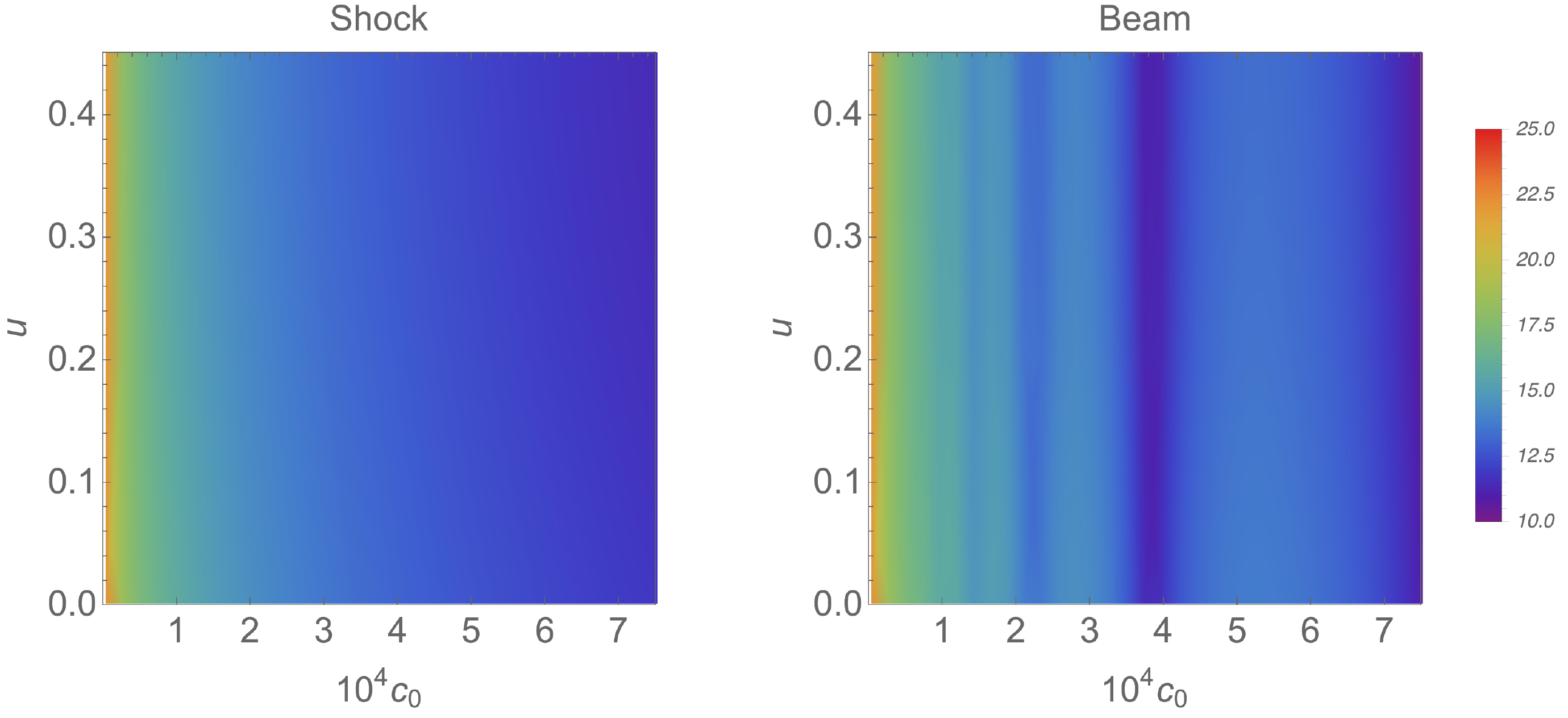}
    \caption{\label{FIG:BWX1} Density plot (log colour scale) of the differential cross section (\ref{BW2PLOT}) at fixed $\xi=4$, as a function of the transverse momentum transfer $c_0 := |c|/m$ and pair longitudinal momentum $u=n\cdot p/n\cdot \ell$. We compare the shockwave and beam backgrounds, for the latter taking $r_0 = 0.01\mu \text{m}$. The oscillations visible in Fig.~\ref{FIG:XSEKT} are again visible here. }
\end{figure}

\begin{figure}[t!]
    \includegraphics[width=0.7\textwidth]{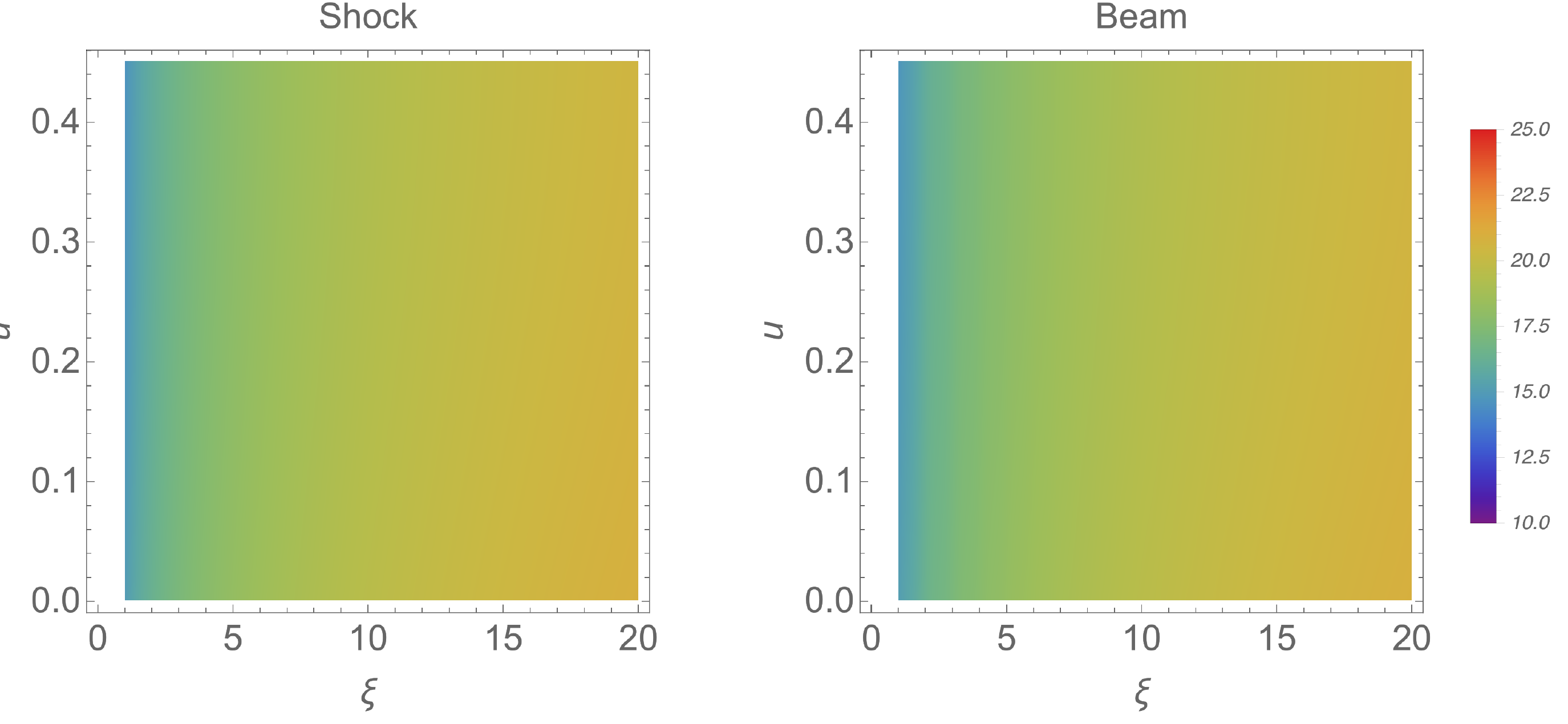}
    \caption{\label{FIG:BWX2} Density plot (log colour scale) of the differential cross section (\ref{BW2PLOT}) at fixed momentum transfer $|c| r_0=1$, as a function of the field strength $\xi$ pair longitudinal momentum $u=n\cdot p/n\cdot \ell$. We compare the shockwave and beam backgrounds, but there is little difference between them.}
\end{figure}

As a result, the cross-section of nonlinear Breit-Wheeler in the ultraboosted beam may also be expressed as a sum over the probabilities for nonlinear Breit-Wheeler in impulsive plane waves, as a direct calculation confirms:
\begin{align}\label{eqn:BWProbability}
    \sigma
    =
    \int \frac{\ud^{2} c_{\lcperp}}{(2\pi)^{2}}
    |W(c)|^{2}
    \bbP_{\text{nbw}}
    \,.
\end{align}
The plane wave probabilities $\bbP_{\text{nbw}}$ are expressed as follows~\cite{Ilderton:2019vot}. 
Define
\begin{align}\label{eqn:H}
    u = \frac{p_{\lcp}}{\ell_{\lcp}} \;, \qquad \qquad  h(u)
    =
    \frac{1}{2}
    -
    \frac{1}{4 u (1 - u)}
    \,.
\end{align}
Recalling the definition of $\pi_{\mu}$ from \eqref{eqn:DressedSpin}, and again writing $c_0:=|c|/m$, the nonlinear Briet-Wheeler probability in an impulsive plane wave is
\begin{align}\label{eqn:BWImpulseProbability}
    \nonumber \bbP_{\text{nbw}}
    &=
    \frac{\alpha m^{2}}{4 \pi^{2}}
    \int \ud^{2} p_{\lcperp}
    \int_{0}^{1} \frac{\ud u}{u} (1 - u) \;
    \bigg[
        \frac{1}{(\ell\cdot \pi)^{2}}
        +
        \frac{1}{(\ell\cdot p)^{2}}
        -
        \frac{2}{(\ell\cdot \pi) (\ell\cdot p)}
        \Big(
            1
            +
            \frac{|c_{\lcperp}|^{2}}{m^{2}}
            h(u)
        \Big) 
    \bigg]
    \\
    &=
    \frac{\alpha}{3 \pi}
    +
    \frac{4 \alpha}{3 \pi}
    \frac{c_{0}^{2} - 1}{c_{0} \sqrt{c_{0}^{2} + 4}}
    \tanh^{-1}\bigg(
        \frac{c_{0}}{\sqrt{c_{0}^{2} + 4}}
    \bigg) 
    \,.
\end{align}
In contrast to nonlinear Compton scattering \eqref{eqn:NLCImpulseProbabilityIntegrated}, the Breit-Wheeler probability is IR finite. 
As a result, the cross section for nonlinear Breit-Wheeler in the shockwave background shows only a single logarithmic behaviour for small momentum transfer. 
In Figures~\ref{FIG:BWX1} and~\ref{FIG:BWX2} we plot the differential cross section obtained  by performing the $p_\LCperp$ integrals in (\ref{eqn:BWImpulseProbability}) but not the $u$-integral: 
\be\label{BW2PLOT}
\frac{1}{r_0^4}\frac{\ud^3\sigma}{\ud u \ud^2 c} = \frac{2\alpha}{\pi r_0^4}|W(c)|^2 u (1-u) \bigg(1-\frac{4 (1+c_0^2 h(u))}{c_0 \sqrt{c_0^2+4}}\tanh^{-1}\bigg[\frac{c_0}{\sqrt{c_0^2+4}}\bigg]\bigg) \;,
\ee
The majority of the structure is seen at momentum transfer of order $c_0 \sim 1/(mr_0)$ which will be small for experimentally relevant beams in which the beam radius must be very much larger than the Compton scale. 
Independent of whether we fix the field strength $\xi$ or the momentum transfer $c_0$, there is only a weak dependence on the longitudinal momentum $u$ of the produced pair.

\subsection{Strong field behaviour}
%
A constant crossed field (CCF) is a plane wave for which $\bf E$ and $\bf B$ are constant and homogeneous. 
It has been conjectured that loop corrections in the Furry expansion, in a CCF, scale at high field strength not with powers of $\alpha$, but with powers of $\alpha\chi^{2/3}$, in which $\chi$ is essentially the product of field strength and probe particle energy~\cite{Ritus1,Narozhnyi:1980dc}. 
This has recently been confirmed for a class of electron self-energy diagrams to all loop orders~\cite{Mironov:2020gbi} (and investigations continue into other loop corrections~\cite{DiPiazza:2020kze,Mironov:2021bmp}).

A consequence of this behaviour is that high external field strengths eventually force the resummation of the loop expansion. 
As such, the `Ritus-Narozhny' (RN) conjecture above is an example of a more general phenomenon~\cite{Heinzl:2021mji}, namely the breakdown and necessary resummation of the Furry expansion, of which there exist many examples in both classical and quantum electrodynamics~\cite{Karbstein:2019wmj,Heinzl:2021mji,Torgrimsson:2021wcj}. 
The quantum physics of the regime in which the Furry expansion requires resummation (and QED becomes in some sense fully non-perturbative) is still hard to access. 
Progress in the classical limit~\cite{Heinzl:2021mji,Torgrimsson:2021wcj} suggests that the regime is one of strong radiation reaction effects, which can lead to novel particle dynamics including chaos, attractors and trapping; for a review see~\cite{Gonoskov:2021hwf}.

It has long been argued that any field will appear to a relativistic particle as (locally) constant and crossed~\cite{Ritus:1985}, and so the `locally constant crossed field approximation' (LCFA) is very commonly used to extend plane wave results to more general fields, including those of beam-beam collisions~\cite{DelGaudio:2018lfm,Yakimenko:2018kih}. 
Now, the fields of the beam considered here are indeed crossed, or null ($F^2 = F{\tilde F} = 0$) as a result of the ultra-boost, but they are also ultra-short: scattering in these fields is constructed from (without approximation) results in plane-waves of ultra-short duration, for which the LCFA fails~\cite{Ilderton:2019vot}. 
That the LCFA is not relevant for the particle-beam collisions described here is further emphasised by the dominance of infra-red effects in the cross-sections presented above: the CCF and LCFA approximations fail completely to describe infra-red physics~\cite{Dinu:2012tj}, see also~\cite{Harvey:2014qla,DiPiazza:2017raw,Ilderton:2018nws}.

Given that the background fields considered here are sourced by ultra-boosted particles, our results are consistent with~\cite{Podszus:2018hnz,Ilderton:2019kqp}, which found that plane wave amplitudes in the high-energy limit~\cite{DiPiazza:2021rum} show a logarithmic dependence on $\chi$, or energy, familiar from QED in vacuum, rather than the $\chi^{2/3}$ power law behaviour of the RN conjecture. 
While the material presented here is only a first investigation into beam-beam collisions, it emphasises that there is more to be understood in the interplay of high field strength and high energy, with regards to the RN conjecture. 
It would therefore be interesting to study loop corrections to the processes considered here; while the behaviour predicted by the RN conjecture does not occur, some other interesting strong-field behaviour could arise.

\section{Summary \label{sec:Summary}}

Motivated by the interest in using beam-beam collisions to access the strong field regime of QED, we have studied the scattering of probe particles from ultraboosted bunches of charge. 
This may be formulated as a background field problem, in which the probe interacts with the boosted collective Coulomb fields of the bunch. 
Such fields are a generalisation of the shockwave background used to describe eikonal, high-energy, scattering.

We solved for both the classical orbit and quantum wavefunctions describing electrons and positrons crossing the generalised shockwave. 
Notably, we found that these wavefunctions are closely related to those on an impulsive (ultra-short) plane wave background. 
This allowed us to construct, without approximation, scattering amplitudes in the ultraboosted beam from amplitudes in impulsive plane waves. 
In effect, the beam seems to act as a stochastic plane wave, and to construct a scattering amplitude one simply averages over amplitudes in plane waves, with an appropriate weight.

We have analysed elastic scattering, nonlinear Compton scattering (photon emission from an electron in the beam) and nonlinear Breit-Wheeler (electron-positron production from a photon in the beam) at tree-level. 
We found that due to the long-rang Coulomb fields of the boosted beam, cross-sections are dominated by infra-red effects. 
We also showed that for high charge density in the beam, meaning strong fields, one can ignore the finite beam radius and model the beam as a single charged particle, or shockwave. For the latter case, all integrals could be performed exactly to obtain cross sections which are dominated by infra-red effects.

In contrast to other approaches, our results are not based on the locally constant field approximation (LCFA). 
Indeed that approximation fails to describe the system considered here. 
This is highlighted by: i.) the importance of infra-red effects in the analysed cross-sections, and ii.) the scaling of amplitudes with high field strength not matching the prediction of the Ritus-Narozhny conjecture. 

There are several avenues for future study. 
One is simply the calculation of other processes of possible experimental interest, such as trident pair production at tree level, and vacuum birefringence at one-loop. 
Higher-order, and ideally all-order, loop corrections to the processes considered here should also be calculated; progress may be made possible due to the ultra-short duration of the background.  
As part of this investigation, one would necessarily study loop corrections in impulsive plane waves which, it is interesting to note, find their direct physical relevance through the boosted beams considered here. 
More generally, the link between impulsive plane waves and shockwaves suggests that recent results on higher-point scattering processes in backgrounds~\cite{Adamo:2020syc,Adamo:2020yzi,Adamo:2021hno,Edwards:2021vhg} or coherent states~\cite{Cristofoli:2021vyo} could be of interest in the context of particle-beam or beam-beam scattering. \\

\emph{We thank Tom Heinzl for providing useful references. The authors are supported by a Royal Society University Research Fellowship (TA), the Leverhulme Trust through RPG-2020-386 (TA) and RPG-2019-148 (AI), and EPSRC through EP/S010319/1 (AI \& AM).}

\end{document}